\documentclass[conference]{IEEEtran}
\IEEEoverridecommandlockouts
\usepackage{cite}
\usepackage{amsmath,amssymb,amsfonts}
\usepackage{algorithmic}
\usepackage{graphicx}
\usepackage{textcomp}
\usepackage{xcolor}
\usepackage{tikz}
\usepackage{textcomp}
\def\BibTeX{{\rm B\kern-.05em{\sc i\kern-.025em b}\kern-.08em
    T\kern-.1667em\lower.7ex\hbox{E}\kern-.125emX}}

    \newcommand\copyrighttext{%
  \footnotesize \textcopyright 2021 IEEE. Personal use of this material is permitted.  Permission from IEEE must be obtained for all other uses, in any current or future media, including reprinting/republishing this material for advertising or promotional purposes, creating new collective works, for resale or redistribution to servers or lists, or reuse of any copyrighted component of this work in other works.
  DOI: {10.23919/IEEECONF54431.2021.9598409}}
\newcommand\copyrightnotice{%
\begin{tikzpicture}[remember picture,overlay]
\node[anchor=south,yshift=10pt] at (current page.south) {\fbox{\parbox{\dimexpr\textwidth-\fboxsep-\fboxrule\relax}{\copyrighttext}}};
\end{tikzpicture}%
}

\begin{document}

\title{Cooperative RADAR Sensors for the Digital Test Field A9 (KoRA9) – Algorithmic Recap and Lessons Learned
\thanks{This work was funded by the German Federal Ministry of Transport and Digital Infrastructure within the project KoRA9 (grant No. 16AVF1032A).}
}

\author{\IEEEauthorblockN{Sören Kohnert, Julian Stähler, Reinhard Stolle}
\IEEEauthorblockA{\textit{Augsburg, University of Applied Science}\\
Augsburg, Germany \\
soeren.kohnert@hs-augsburg.de}
\and
\IEEEauthorblockN{Florian Geissler}
\IEEEauthorblockA{\textit{Dependability Research Lab, Intel Labs} \\
Munich, Germany \\
florian.geissler@intel.com}
}

\maketitle
\copyrightnotice
\begin{abstract}
Infrastructure sensing systems in combination with
Infrastructure-to-Vehicle communication can be used to enhance sensor data obtained from the perspective of a vehicle, only.
This paper presents a system consisting of a radar sensor network installed
at the side of the street, together with an Edge Processing
Unit to fuse the data of different sensors. 
Measurements taken by the demonstrator are shown, the system
architecture is discussed, and some lessons learned are presented. 
\end{abstract}


\section{Introduction}
Automated vehicles will play a major role in the traffic of our future. Today, production vehicles are already driving automated at speeds of up to 60~km/h on the motorways \cite{vw_2020}. It is difficult to reach higher speeds, in part because the vehicle’s perception is lacking the foresight of a human driver. Additional roadside sensor networks operating as perception systems can not only contribute to foresight, but also add functional safety and consequently reach higher levels of automotive safety integrity (ASIL) \cite{iso26262}. Use cases can go as far as the transmission of emergency trajectories in the case of a fault in the vehicle’s own perception or location system \cite{pechinger_2020}. 

A first research project that came up with the idea of infrastructure assisting traffic, was the PATH \cite{Shladover_1992} project. More recently, Providentia \cite{kraemmer_2020} monitored a highway using a combination of radar and camera mounted on a gantry. The installation on a gantry has the advantages of smaller influence on occlusion as well as a better aspect angle of illumination for the radar sensor. For the validation process, only cameras were used while the radar sensors were disabled as they were found to reduce the overall data quality. 

The goals of this project were to adapt automotive radar sensors for infrastructure use and to investigate the potential of such a foresight and functional safety system along the motorway. 

In this contribution, we present design considerations for such a sensor network, taking typical restrictions for a motorway use case into account. We demonstrate the full functionality of system architecture, from the RADAR sensors to the Edge Processing Unit, and focus on the algorithmic level to discuss the results and lessons learned during the project.
We will present qualitative results from simulations of sensor coexistence interference scenarios for such sensor networks. Furthermore, we will discuss restrictions of and thoughts about the sensor placement along the motorway, considering sensor coverage as well as the occlusion of vehicles by other vehicles.

\section{System Architecture}
\begin{figure*}[ht]
 \centering
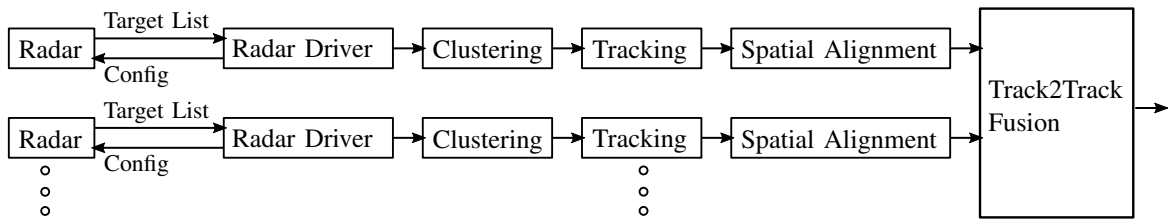
\caption{System architecture of demonstrator of the KoRA9 projects. The data flow is from left to right. Only the processing steps up to the target list generation is done on the sensor. Everything beyond that is processed on the edge computer. The result of the T2T fusion is communicated to the traffic participants.}
\label{fig:system_architecture}
\end{figure*}

The overall system architecture is shown in Fig.~\ref{fig:system_architecture}.
Frequency-modulated continuous-wave (FMCW) RADAR systems operating at 77 GHz are mounted at equal distances along the roadside. 
The architecture of the radar signal processing chip allows a hardware implementation of the signal processing up to the target list. This target list is transmitted over Ethernet to the Edge Processing Unit. A driver brings the sensor data in the Robot Operating System (ROS) \cite{ros} domain where all further processing is done. For clustering, a DBSCAN algorithm is used in the range Doppler domain to group the targets to objects. A linear Kalman filter is used for tracking. The data get spatially aligned before the Track2Track (T2T) fusion. There, a fusion approach similar to \cite{Houenou_2012} is adopted.

\subsection{Sensors}
\begin{figure}[ht]
\includegraphics[trim=0 0 0 0, clip, width=82mm]{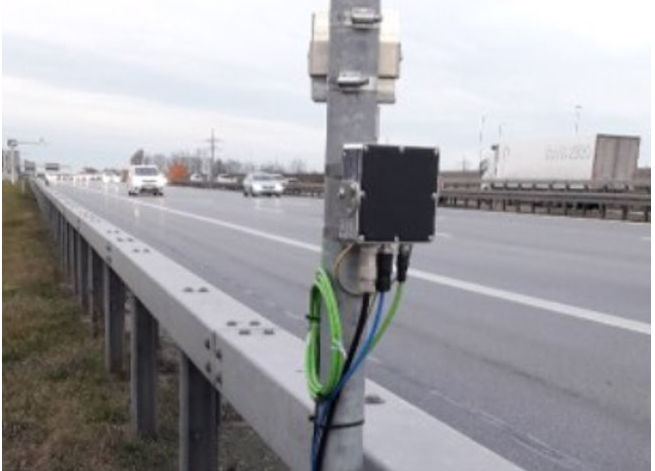}
\caption{Two radar sensor prototypes, installed at the demonstrator, looking in the direction of oncoming and departing traffic, respectively.}
\label{fig:sensor_a9}
\end{figure}

Based on Infineon`s 77$\,$GHz \textit{RXS8160PL} radar sensor IC and \textit{Aurix 2G TC3xx} controller, a sensor for infrastructure perception system use on motorways was developed. Some key parameters of the sensor and its configuration are shown in Table~\ref{tab:radPara}. The sensor itself is shown in Fig.~\ref{fig:sensor_a9}. It is optimised for an unambiguous velocity measurement and the desired range that came from the learnings of the completeness analysis described in Section~\ref{section:occlusion}.


\renewcommand{\arraystretch}{1.1}
\begin{table}[ht]
	\caption{Main parameters of the FMCW radar.}
	\small
	\centering
	\begin{tabular}{l|l|l|l}
		Parameter & Symbol & Value & Unit \\ \hline
		Centre Frequency & $f_0$ & 76.5 & GHz \\
		Bandwidth & $BW$ & 150 & MHz \\
		Sampling Frequency & $f_s$ & 12.5 & MHz \\
		Chirp Duration & $T_R$ & 20.48 & $\mathrm{\mu}$s \\
		Unambiguous Range & $R_{max}$ & 127.2 & m \\
		Unambiguous Speed & $v_{max}$ & 66.56 & m/s\\
		Measurement Cycle Rate & $f_m$ & 20 & Hz\\
	\end{tabular}
\label{tab:radPara}
\end{table}
\renewcommand{\arraystretch}{1.0}

\begin{figure*}[ht]
\includegraphics[trim=112 60 20 80, clip, width=200mm]{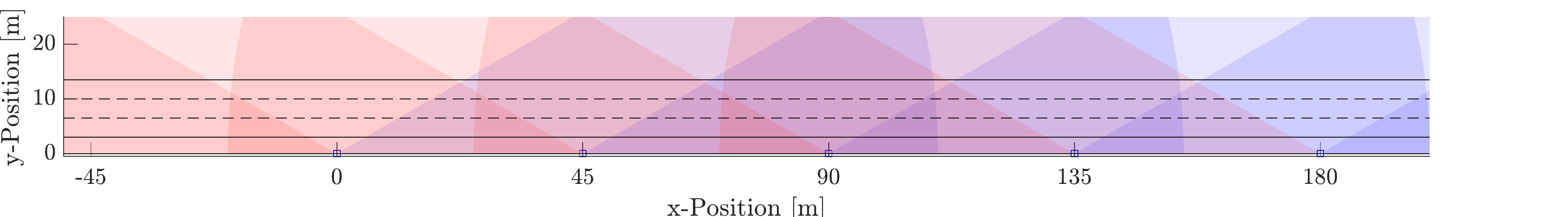}
\caption{Simplified model of the surveyed motorway section. The poles are placed along the road every 45$\,$m (blue squares). Each pole carries two sensors. One looking towards the oncoming (Field of Views (FoVs) in light red) and one looking towards the departing traffic (FoVs in light blue). This design is the result of an optimization under the given constraints as done in \cite{geissler_2019}.}
\label{fig:kora9Layout}
\end{figure*}

\subsection{Fusion}
For the inter-sensor fusion, a high-level fusion concept similar to Honenou et al. \cite{Houenou_2012}  was chosen for the sake of simplicity and reusability. The same fusion algorithm can be deployed in the vehicle, fusing the data generated in the infrastructure in the vehicle once more with the vehicle’s own environment recognition. The fusion operator weighs the object states with respect to their inverse (co-)variance. It can handle out-of-sequence measurements and can incorporate the history of the tracks during the association step. 

\subsection{Demonstrator}
The demonstrator was installed at the motorway A9 close the the town of Neufahrn, north of Munich. The location was chosen due to the high traffic throughput and a side lane that can be opened to manage temporarily increased traffic. The demonstrator consists of 10 sensors distributed across 5 poles. Each pole holds two sensors, looking in the direction of oncoming and departing traffic at an angle of  $\pm 15\deg$ away from the roadside to the middle, respectively. The layout is displayed in Fig.~\ref{fig:kora9Layout}.

\begin{figure*}[ht]
 \centering
\includegraphics[trim=111 82 20 80, clip, width=200mm]{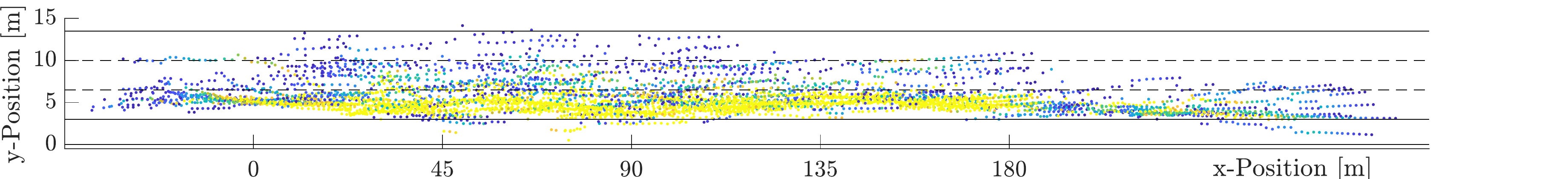}
\caption{Estimates of object positions of $2\,\mathrm{min}$ of motorway traffic. The plausibility estimates are color coded. The brighter the color, the higher the plausibility.
The plausibility is higher in areas with higher sensor coverage, as well as on the first lane.}
\label{fig:measurements_a9}
\end{figure*}

Fig. \ref{fig:measurements_a9} shows the plausibility score of a 2-minutes-long measurement with the demonstrator. The brighter the color the higher the plausibility of this track.
While having a low score at the beginning and end of the demonstrator’s FoV, it is high on the first lane between $0$ and $180\,\mathrm{m}$, where the sensors provide the best street coverage.
The consistently high plausibility in this area indicates a functioning fusion. Otherwise we would see such initialisation phases in the beginning and end of every FoV of a sensor, not only once in the beginning and end of the FoV of the whole demonstrator.

\section{Lessons Learned}
Throughout the project, the project team faced multiple challenges. Some were solved during the project, some are to be solved in future projects. This section summarises these lessons learned.

\subsection{Where to install sensors}
\label{section:occlusion}
The sensor network parameters like beam width and distance between sensors are found optimizing for the completeness of information gained on the test field under consideration of possible occlusions caused by vehicles.
The input of the optimisation is real traffic data acquired on the same road where the demonstrator is located \cite{geissler_2018}. 
The installation of the sensor on the right side results in information loss, as passing trucks often occlude other objects. Nevertheless, right side installment has other advantages as e.g. facilitating maintenance work. \cite{geissler_2019} 

Radar test measurements using a truck with semitrailer and vehicles were performed to get the optimal height. The lower the height, the greater the probability that the sensor beam passes underneath vehicles, especially trucks. 
On the other hand, heavy snowfall in winter can be an issue. When clearing snow, the snow piles up at the edge of the road covering potentially low mounted sensors. High mounted sensors might be a safety hazard, becoming a projectile in case of collision. To avoid occlusion from the guard rail, an installation height of $1.3\,\mathrm{m}$ was finally chosen.

\subsection{Aspect Angle of Illumination}
\label{section:aspect_angle}
\begin{figure}[ht]
\includegraphics[trim=20 2 35 5, clip, width=82mm]{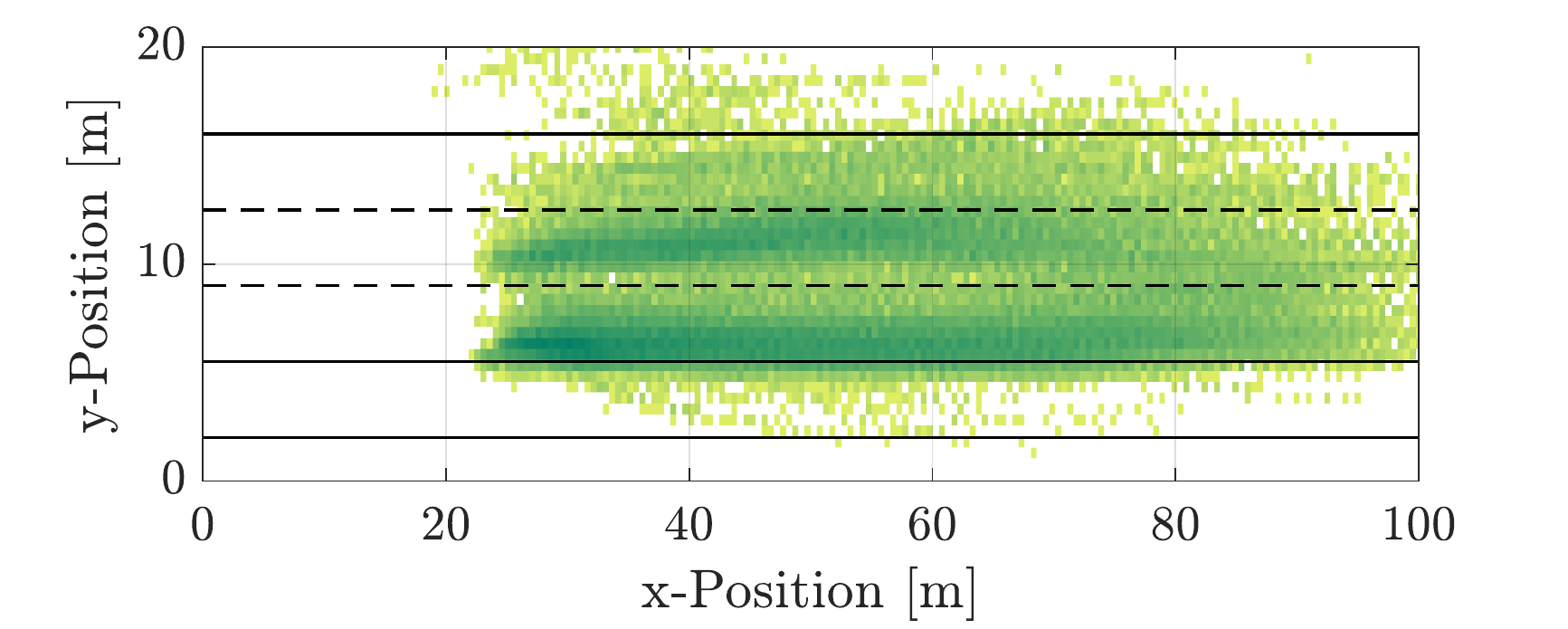}
\caption{Heat map of objects detected over a $30\,\mathrm{min}$ period. The sensor is located in the origin and looking along the departing traffic.}
\label{fig:rad9_objects}
\end{figure}

Fig.~\ref{fig:rad9_objects} shows a heat map of the object positions generated by one sensor that is directed in the departing traffic. The darker the green, the more objects are detected in this area. For better visualisation, the logarithm of the amount of vehicles is shown. The emergency lane was not open to traffic while recording. Whereas the detection works well for the first lane, almost no objects got detected on the third lane. We expect to detect less vehicles on that lane, due to the obligation to drive on the right-hand side of the road and occlusion by other vehicles, but still would have expected a visible line indicating the travel direction of the vehicles. Geary et al. \cite{geary} found that depending on the aspect angle of illumination, the radar cross section (RCS) can differ up to $30\,\mathrm{dB}$. Therefore, unfavorable illumination angles on the most distant lane may be another reason for the observed lowered detection rates. 

\subsection{Tracking}
The key challenge in tracking is the velocity of the objects and the consequently short time that the vehicles spend in the sensor's Field of View (FoV), typically \textless $\,2\,\mathrm{s}$. We found that a Kalman Filter with a simple constant velocity motion model works best for tracking since it converges faster than more complex models. A crucial part is the track initialisation. Since the sensors are installed at fixed positions and face the motorway at fixed angles, we initialised the tracks already with the assumed driving direction and an assumed speed of 80$\,$km/h. The assumed speed is taken to be the center of the observed speed value range, which is given by 0 and 160$\,$km/h. Moreover, it is possible to correct the measured radial velocity by the assumed driving direction.
For a next project we suggest to use a lower level fusion approach for a sensor setup with only one kind of sensor. 
This way, the tracks do not have to be reinitialised for each radar FoV. That would concentrate the processing to a single tracking node, saving overall computational effort, but at the risk of creating a bottleneck in the fusion node. A discussion of the pros and cons of different fusion concepts can be found here \cite{aeberhardt}.

Another challenge is the influence of the point target assumption on the tracking and fusion of long objects like trucks.  Assuming that targets are not spatially extended is problematic, especially when sensors are mounted in such a way that they can perceive opposite sides of the object. One sensor will see the front of the vehicle and another one the back, not being able to estimate the length. A truck with semitrailer combination is typically about 14$\,$m long, but can also reach up to 20$\,$m for trucks with trailer combinations. Still, the objects are unnaturally close for two independent objects and can therefore be merged to a single long object, if they are estimated to be on the same lane.

Depending on the use case it can make sense to sacrifice degrees of freedom in the tracking and therefore simplify it. This can be achieved by e.g. removing the degree of heading and restricting the vehicles to the different lanes. 

\subsection{Sensor Coexistence Interference}

\begin{figure}[ht]
\centering
\begingroup%
  \makeatletter%
  \providecommand\color[2][]{%
    \errmessage{(Inkscape) Color is used for the text in Inkscape, but the package 'color.sty' is not loaded}%
    \renewcommand\color[2][]{}%
  }%
  \providecommand\transparent[1]{%
    \errmessage{(Inkscape) Transparency is used (non-zero) for the text in Inkscape, but the package 'transparent.sty' is not loaded}%
    \renewcommand\transparent[1]{}%
  }%
  \providecommand\rotatebox[2]{#2}%
  \newcommand*\fsize{\dimexpr\f@size pt\relax}%
  \newcommand*\lineheight[1]{\fontsize{\fsize}{#1\fsize}\selectfont}%
  \ifx\svgwidth\undefined%
    \setlength{\unitlength}{128.29387189bp}%
    \ifx\svgscale\undefined%
      \relax%
    \else%
      \setlength{\unitlength}{\unitlength * \real{\svgscale}}%
    \fi%
  \else%
    \setlength{\unitlength}{\svgwidth}%
  \fi%
  \global\let\svgwidth\undefined%
  \global\let\svgscale\undefined%
  \makeatother%
  \begin{picture}(1,0.5381313)%
    \lineheight{1}%
    \setlength\tabcolsep{0pt}%
    \put(0,0){\includegraphics[width=\unitlength,page=1]{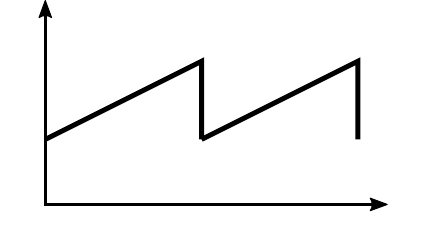}}%
    \put(0.1395883,0.47425284){\color[rgb]{0,0,0}\makebox(0,0)[lt]{\lineheight{1.25}\smash{\begin{tabular}[t]{l}$T_\mathrm{sweep}$\end{tabular}}}}%
    \put(0.86247166,0.2918219){\color[rgb]{0,0,0}\makebox(0,0)[lt]{\lineheight{1.25}\smash{\begin{tabular}[t]{l}BW\end{tabular}}}}%
    \put(0,0){\includegraphics[width=\unitlength,page=2]{ramp.pdf}}%
    \put(0.37460055,0.00117224){\color[rgb]{0,0,0}\makebox(0,0)[lt]{\lineheight{1.25}\smash{\begin{tabular}[t]{l}Time\end{tabular}}}}%
    \put(0.05676969,0.1378358){\color[rgb]{0,0,0}\rotatebox{90}{\makebox(0,0)[lt]{\lineheight{1.25}\smash{\begin{tabular}[t]{l}Frequency\end{tabular}}}}}%
    \put(0,0){\includegraphics[width=\unitlength,page=3]{ramp.pdf}}%
    \put(0.26737846,0.13203072){\color[rgb]{0,0,0}\makebox(0,0)[lt]{\lineheight{1.25}\smash{\begin{tabular}[t]{l}No. of ramps\end{tabular}}}}%
    \put(-0.80438874,1.42371013){\color[rgb]{0,0,0}\makebox(0,0)[lt]{\begin{minipage}{2.74759808\unitlength}\raggedright \end{minipage}}}%
  \end{picture}%
\endgroup%

\caption{FMCW ramp parameters of interest for the presented interference analysis. }
\label{fig:ramp}
\end{figure}

Having such a dense sensor network will cause a lot of sensor coexistence interference among sensors \cite{brooker}. An analysis was done to to provide typical interference patterns in the range Doppler domain. 
For this purpose, the response of a single radar target at intermediate frequency level was simulated, the target being illuminated by the radar itself (victim) and by a second radar transmitter (interferer), operating at different measurement bandwidths and sweep times. The investigated scenarios are shown in Table~\ref{tab:interference}. Fig.~\ref{fig:ramp} shows the definition of the parameters for the example of two consecutive FMCW ramps. We investigate the influence of three parameters: The sweep bandwidth (BW), the sweep time ($T_\mathrm{sweep}$), and the number (No.) of ramps that overlap between both sensor's signals. 

\renewcommand{\arraystretch}{1.1}
\begin{table}[ht]
	\caption{Qualitative interference analysis. The numbers represent the ratio between the ramps of the interfering sensor and the victim sensor e.g. $1.01 = BW_{interferer} / BW_{victim}$ }
	\small
	\centering
	\begin{tabular}{c|c|c|c}
		Scenario & BW & $T_\mathrm{sweep}$ & No. of ramps  \\ \hline
		(1) & $1$ & $1$ & $1$ \\
		(2) & $1.01$ & $1$ & $1$ \\
		(3) & $1.2$ & $1$ & $1$ \\
		(4) & $1.01$ & $1$ & $10/32$ \\
		(5) & $1$ & $1.1$ & $1$ \\
		(6) & $1$ & $1.01$ & $3/32$ \\
	\end{tabular}
\label{tab:interference}
\end{table}
\renewcommand{\arraystretch}{1.0}

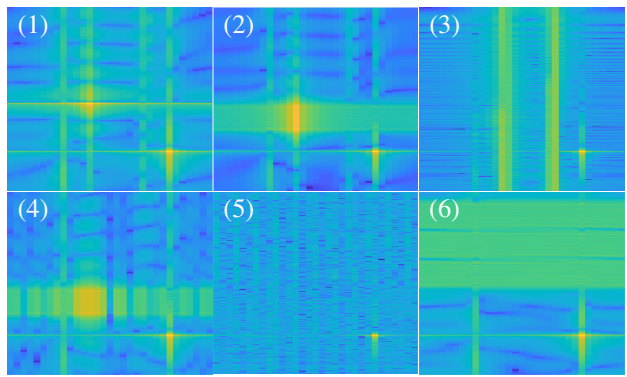
\begin{figure}[ht]
\begingroup%
  \makeatletter%
  \providecommand\color[2][]{%
    \errmessage{(Inkscape) Color is used for the text in Inkscape, but the package 'color.sty' is not loaded}%
    \renewcommand\color[2][]{}%
  }%
  \providecommand\transparent[1]{%
    \errmessage{(Inkscape) Transparency is used (non-zero) for the text in Inkscape, but the package 'transparent.sty' is not loaded}%
    \renewcommand\transparent[1]{}%
  }%
  \providecommand\rotatebox[2]{#2}%
  \newcommand*\fsize{\dimexpr\f@size pt\relax}%
  \newcommand*\lineheight[1]{\fontsize{\fsize}{#1\fsize}\selectfont}%
  \ifx\svgwidth\undefined%
    \setlength{\unitlength}{232.44094329bp}%
    \ifx\svgscale\undefined%
      \relax%
    \else%
      \setlength{\unitlength}{\unitlength * \real{\svgscale}}%
    \fi%
  \else%
    \setlength{\unitlength}{\svgwidth}%
  \fi%
  \global\let\svgwidth\undefined%
  \global\let\svgscale\undefined%
  \makeatother%
  \begin{picture}(1,0.59475469)%
    \lineheight{1}%
    \setlength\tabcolsep{0pt}%
    \put(0,0){\includegraphics[width=\unitlength,page=1]{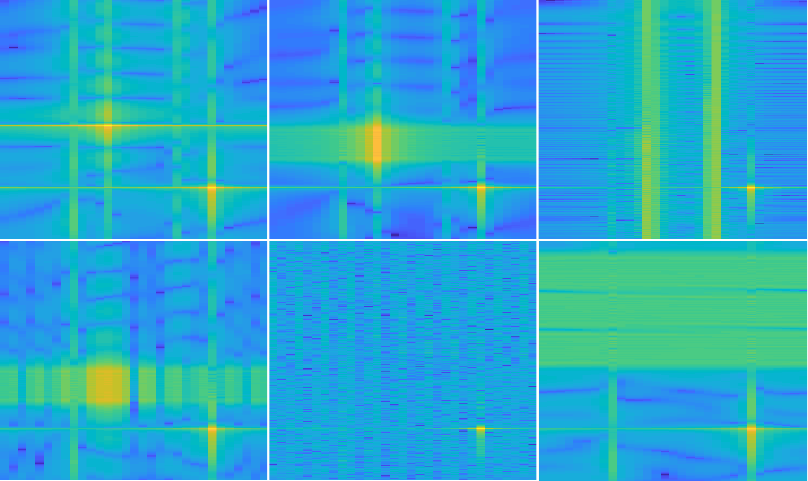}}%
    \put(0.01632368,0.55417245){\color[rgb]{0,0,0}\makebox(0,0)[lt]{\lineheight{1.25}\smash{\begin{tabular}[t]{l}\textcolor{white}{(1)}\end{tabular}}}}%
    \put(0.3497675,0.55421767){\color[rgb]{0,0,0}\makebox(0,0)[lt]{\lineheight{1.25}\smash{\begin{tabular}[t]{l}\textcolor{white}{(2)}\end{tabular}}}}%
    \put(0.68357328,0.55470993){\color[rgb]{0,0,0}\makebox(0,0)[lt]{\lineheight{1.25}\smash{\begin{tabular}[t]{l}\textcolor{white}{(3)}\end{tabular}}}}%
    \put(0.01619327,0.25571313){\color[rgb]{0,0,0}\makebox(0,0)[lt]{\lineheight{1.25}\smash{\begin{tabular}[t]{l}\textcolor{white}{(4)}\end{tabular}}}}%
    \put(0.34972295,0.25585472){\color[rgb]{0,0,0}\makebox(0,0)[lt]{\lineheight{1.25}\smash{\begin{tabular}[t]{l}\textcolor{white}{(5)}\end{tabular}}}}%
    \put(0.68327541,0.25585472){\color[rgb]{0,0,0}\makebox(0,0)[lt]{\lineheight{1.25}\smash{\begin{tabular}[t]{l}\textcolor{white}{(6)}\end{tabular}}}}%
  \end{picture}%
\endgroup%

\caption{Range Doppler plot of six interference scenarios. The target on the left is the target produced by the interfering sensor, whereas the target on the lower right is a simulated target of the victim sensor. The range dimension is on the horizontal axis, whereas the Doppler dimension is on the vertical axis.}
\label{fig:RDPlot}
\end{figure}

Fig.~\ref{fig:RDPlot} shows the range Doppler plots for typical interference scenarios. Most dangerous are scenarios (1), (2), and (4). The more similar the ramps are, the more does the ghost target resemble a real target. The ghost target smears in both the range and Doppler domain. (5) is probably the most common interference. If the ramp of the interfering sensor sweeps through the IF band of the victim sensor, an increased noise floor results. 
The anomalies that have been found can be used to develop algorithms to detect and mitigate the effect of sensor coexistence interference.

\section{Conclusion}
A full scale demonstrator for the use of radar sensors to monitor motorways in quasi-real-time was presented. The demonstrator showed that such systems can detect, process, and transmit an environmental model of road sections to the vehicles to extend and verify the on-board perception of this vehicles. Even though limitations in the radar sensor development process prevented the sensor system from achieving full street coverage by now, our data indicates that utilizing fully developed sensors will overcome these limitations. Apart from the sensors the the demonstrator was able to demonstrate the whole system, from detection, to processing and the transmission to the vehicles.
Efforts will probably shift from a rather deterministic motorway scenario to inner city scenarios with challenges like the safety for vulnerable road users.

\section{Acknowledgements}
Thanks to Dominik Zoeke, Fabian Kurz (both Siemens Mobility GmbH) and Simon Achatz (Infineon AG) for laying the foundations for the project specific algorithm development.

\medskip

\bibliographystyle{./bibliography/IEEEtran}
\bibliography{./bibliography/IEEEexample}

\vspace{12pt}

\end{document}